\documentclass[aps,prb,showpacs,amsmath]{revtex4}

\usepackage{graphicx}

\begin{document}

\title{
Polaron action for multimode dispersive phonon systems
}

\author{P.\ E.\ Kornilovitch}
\email{pavel.kornilovich@hp.com}
\affiliation{
Hewlett-Packard Company, Corvallis, Oregon, 97330, USA
}

\date{\today}

\begin{abstract}

Path-integral approach to the tight-binding polaron is extended to multiple optical 
phonon modes of arbitrary dispersion and polarization.  The non-linear lattice effects
are neglected.  Only one electron band is considered.  The electron-phonon interaction
is of the density-displacement type, but can be of arbitrary spatial range and shape. 
Feynman's analytical integration of ion trajectories is performed by transforming the 
electron-ion forces to the basis in which the phonon dynamical matrix is diagonal.         
The resulting polaron action is derived for the periodic and shifted boundary 
conditions in imaginary time.  The former can be used for calculating polaron
thermodynamics while the latter for the polaron mass and spectrum.  The developed
formalism is the analytical basis for numerical analysis of such models by 
path-integral Monte Carlo methods.  

\end{abstract}


\maketitle

\section{ \label{sec:one}
Introduction
}

In the last two decades, quantum Monte Carlo (MC) simulation methods proved to 
be a powerful theoretical tool in analyzing models with strong electron-phonon 
interactions. The pioneering works by Hirsch and Fradkin on the many-polaron
one-dimensional Holstein model \cite{Hirsch1982} and the Su-Schrieffer-Heeger
polaron \cite{HirschSSH}, by de Raedt and Lagendijk on the Holstein polaron 
\cite{DeRaedt1982} and bipolaron \cite{DeRaedtBipolaron}, and by 
Alexandrou and Rosenfelder on the Fr{\"o}hlich polaron \cite{Alexandrou1992}, 
were followed more recently by the continuous-time path-integral MC 
\cite{Kornilovitch1998}, diagrammatic MC \cite{Prokofev1998,Macridin2004}, and 
the Lang-Firsov MC \cite{Hohenadler2004} methods. The new methods have expanded 
the list of calculable polaron properties, both static and dynamic. However, 
expansion in terms of analysis of more complex polaron models has been slow. The 
bulk of new literature is still devoted to the two canonical polaron models: 
Fr\"ohlich model of the ionic solid \cite{Froehlich} and Holstein model of 
molecular crystal \cite{Holstein1959}. While these models are important, the new 
MC methods are powerful and versatile enough to study others. Notable exceptions 
from this trend are the investigations of the Su-Schrieffer-Heeger model 
\cite{HirschSSH}, Jahn-Teller polaron \cite{Kornilovitch2000}, and of the $tJ$ 
model with electron-phonon interaction \cite{Mishchenko2004}.    

We recently started a series of generalizations of the continuous-time path 
integral MC method away from the standard Holstein interaction with localized 
phonons on cubic lattices. A long-range electron-phonon interaction was 
investigated in Refs.~[\onlinecite{longrange}], and different Bravais lattices
in Ref.~[\onlinecite{Hague2005}]. In this paper, we take the first step toward 
generalizing the method on dispersive and multimode phonon systems. A critical
element of the method is analytical integration over ion trajectories in the 
path-integral expression for the polaron partition function. This leads to a 
polaron action, expressed as a functional over the electron path, which includes 
all the effects of the phonon dynamics and electron-phonon interaction. For the 
Einstein phonons, the integration results in the single-oscillator formula 
derived by Feynman \cite{Feynman1955,Feynman1972}. In more complex cases, the
electron-ion forces must be first transformed to the basis of the lattice normal 
modes, then the normal modes integrated out using the Feynman formula, and then
the final result transformed back to the orignal basis. To our knowledge, the 
only investigation of this kind in the polaron MC literature was done by 
de Raedt and Lagendijk \cite{DeRaedt1984} who studied the one-dimensional 
Holstein polaron with dispersive optical phonons by the discrete-time
path-integral MC.  They found that the transition into a self-trapped state
always takes place for for dispersive phonons as well as for dispersionless 
phonons but the transition point depends on the details of the phonon dispersion.  
Their derivation now needs to be repeated for continuous imaginary time and 
generalized to multiple phonon modes.
Recently, Zoli \cite{Zoli2005} calculated the phonon path integral for the 
Holstein model in the semiclassical approximation where the retarded nature
of the phonon action was neglected. 

Phonon dispersion is expected to affect polaron properties, most notably the 
polaron effective mass and hence mobility.  The forces acting between the 
ions spread out the deformation even if the electron-phonon interaction is
local.  When the electron hops between the lattice cells, partial deformation 
already exists, which increases the phonon overlap integral.  Thus in general,
dispersion {\em reduces} the polaron mass.  In this sense, the dispersion is 
analogous to a long-range electron-phonon interaction \cite{longrange}.

The purpose of the present paper is to carry out general phonon integration to 
enable MC studies of novel polaron models.  The main analytical result is given 
below in equations (\ref{eq:thirtysix})-(\ref{eq:thirtyeight}) and 
(\ref{eq:thirtythree})-(\ref{eq:thirtyfour}).  The expressions are rather complex, 
however, and the practical impact of the formalism will depend on the efficiency 
of numerical methods used to evaluate the polaron action at each MC update. This 
is likely to become the subject of future work.

\section{ \label{sec:two}
Shift partition function
}

Consider the general problem of a lattice electron interacting with an 
arbitrary system of ions via a density-displacement interaction.  In the 
path integral formalism \cite{Feynman1955}, the electron is described by an 
imaginary-time path ${\bf r(\tau)}$, where ${\bf r}$ is the discrete lattice 
coordinate at ``time'' $\tau$, and $0 \leq \tau \leq \beta = (k_B T)^{-1}$.
In this paper, only one Wannier state per init cell is assumed.  The ion 
system is represented by $p \geq 1$ continuous degrees of freedom 
$\xi_{{\bf n}s}$ (displacements) where ${\bf n}$ numbers the unit cells, and 
$1 \leq s \leq p$.  The mass of $s$-th degree of freedom is $m_s$.  Ions 
interact with each other via pairwise {\em force} constants 
$v_{{\bf n}s, {\bf n}' s'}$, which are functions of the difference 
${\bf n}-{\bf n}'$ only.  Cubic and higher terms in the lattice potential 
energy are neglected.  A displacement $\xi_{{\bf n}s}$ is acted 
upon by the electron with force $f_{{\bf n}s}(\tau)= f_{{\bf n}s}[{\bf r}(\tau)]$, 
so that the interaction potential is $-f_{{\bf n}s}(\tau) \xi_{{\bf n}s}$.  
The imaginary time action of such a system is given by
\begin{eqnarray}
A [{\bf r}(\tau), \xi_{{\bf n}s}(\tau)] & = & T[{\bf r}(\tau)] + 
A_{\rm ph}[\xi_{{\bf n}s}(\tau)] + A_{\rm e-ph}[{\bf r}(\tau), \xi_{{\bf n}s}(\tau)]  
\nonumber \\
& = & T[{\bf r}(\tau)] - \!\! 
\sum_{{\bf n}s} \!\! \int^{\beta}_{0} \! \frac{m_s {\dot \xi}^2_{{\bf n}s}(\tau)}{2 \hbar^2} \: 
{\rm d} \tau -   
\frac{1}{2} \!\! \sum_{{\bf n}s,{\bf n}'s'} \! \int^{\beta}_{0} \!\!\! v_{{\bf n}s,{\bf n}'s'} 
\xi_{{\bf n}s} (\tau) \xi_{{\bf n}'s'}(\tau)  \, {\rm d} \tau +
\sum_{{\bf n}s} \int^{\beta}_{0} \!\!\! f_{{\bf n}s}[{\bf r}(\tau)] \xi_{{\bf n}s}(\tau) 
{\rm d} \tau  .
\label{eq:one}
\end{eqnarray}
Here the first term $T$ represents the kinetic energy of the electron.  
It is not the subject of the present work, see instead 
Refs.~[\onlinecite{Kornilovitch1998,Kornilovitch2000}].  Note that the action 
$A$ is dimensionless. The statistical weight of a configuration is given by 
$\exp{(A)}$. The central object of the polaron path-integral quantum Monte Carlo 
method is the {\em shift partition function}
\begin{equation}
Z_{\Delta {\bf r}} = \sum_{{\bf r}_0} \int \prod_{{\bf n}s} {\rm d} \xi_{{\bf n}s0}
\int^{{\bf r}(\beta) = {\bf r}_0 + \Delta {\bf r}; \,
\xi_{{\bf n}s}(\beta) = \xi_{({\bf n}-\Delta {\bf r})s0} }
_{{\bf r}(0) = {\bf r}_0; \, \xi_{{\bf n}s}(0) = \xi_{{\bf n}s0}} 
{\cal D} {\bf r}(\tau) \, {\cal D} \xi_{{\bf n}s}(\tau) \: 
e^{A [ \, {\bf r}(\tau), \, \xi_{{\bf n}s}(\tau)]} \: .
\label{eq:two}
\end{equation}
Here the inner path integral is taken under shifted boundary conditions in imaginary time.
This means that the entire real-space configuration at the end point $\tau = \beta$ (which 
comprises the electron and phonon coordinates) is identical to the configuration at the 
end point $\tau = 0$ {\em up to a parallel shift} by a real space vector $\Delta {\bf r}$.  
The outer ordinary but multidimensional integral is over all possible configurations at 
$\tau = 0$.  

Since the ion coordinates enter the action linearly and quadratically, path integration over 
$\xi_{{\bf n}s}(\tau)$ in (\ref{eq:one}) can be done analytically.  We first calculate a more 
general path integral
\begin{equation}
W [\xi_{{\bf n}s0}, \xi_{{\bf n}s\beta}; {\bf r}(\tau)] = 
\int^{\xi_{{\bf n}s}(\beta) = \xi_{{\bf n}s\beta} }_{\xi_{{\bf n}s}(0) = \xi_{{\bf n}s0}} 
{\cal D} \xi_{{\bf n}s}(\tau) \: 
e^{A_{\rm ph} [\, \xi_{{\bf n}s}(\tau)] + 
A_{\rm e-ph}[\, {\bf r}(\tau), \, \xi_{{\bf n}s}(\tau)]} \: .
\label{eq:three}
\end{equation}
Here, in contrast to eq.~(\ref{eq:two}), the end coordinates of ions are not restricted
in any way. $W$ is a function of the end coordinates $\xi_{{\bf n}s0}$, 
$\xi_{{\bf n}s \beta}$, and a functional of the whole electron path.  
Path integration is performed by a method devised by Feynman \cite{Feynman1972}.  
Variation of the exponent with respect to $\xi_{{\bf n}s}(\tau)$ subjected to fixed 
boundary conditions yields a Euler-Lagrange equation for the stationary path 
${\bar \xi}_{{\bf n}s}(\tau)$
\begin{equation}
\frac{m_s}{\hbar^2} \: {\ddot {\bar \xi}}_{{\bf n}s}(\tau) - 
\sum_{{\bf n}' s'} v_{{\bf n}s,{\bf n}'s'} {\bar \xi}_{{\bf n}'s'}(\tau) + 
f_{{\bf n}s}(\tau) = 0 \: .
\label{eq:four}
\end{equation}
This is a system of $pN$ coupled second order ordinary inhomogeneous differential equations
for functions ${\bar \xi}(\tau)$ with boundary conditions
\begin{eqnarray}
{\bar \xi}_{{\bf n}s}(0) = \xi_{{\bf n}s0} \: ,
\label{eq:five} \\
{\bar \xi}_{{\bf n}s}(\beta) = \xi_{{\bf n}s\beta} \: .
\label{eq:six}
\end{eqnarray}
The inhomogeneous terms $f_{{\bf n}s}(\tau)$ are functionals of the electron path 
${\bf r}(\tau)$ and as such are arbitrary functions of imaginary time $\tau$.  When 
the solution of eq.~(\ref{eq:four}) is found, $W$ is calculated by performing a 
path shift in (\ref{eq:three}): 
\begin{eqnarray}
\xi_{{\bf n}s}(\tau) & = & {\bar \xi}_{{\bf n}s}(\tau) + \eta_{{\bf n}s}(\tau) \: ,
\label{eq:seven} \\
\eta_{{\bf n}s}(0) & = & \eta_{{\bf n}s}(\beta) = 0 \: .
\label{eq:eight}
\end{eqnarray}
Upon substitution of (\ref{eq:seven}) in (\ref{eq:three}) the action 
$A_{\rm ph} + A_{\rm e-ph}$ separates in three terms: 
$A_1$ that depends only on ${\bar \xi}$, $A_2$ that depends only on $\eta$, and the 
mixed term $A_3$ that depends on both ${\bar \xi}$ and $\eta$. The mixed term vanishes 
identically by virtue of the classical equations (\ref{eq:four}).  $W$ becomes
\begin{equation}
W [\xi_{{\bf n}s0}, \xi_{{\bf n}s\beta}; {\bf r}(\tau)] = 
e^{A_1[\, {\bar \xi}_{{\bf n}s}(\tau)]} 
\int^{\eta_{{\bf n}s}(\beta) = 0 }_{\eta_{{\bf n}s}(0) = 0} {\cal D} \eta_{{\bf n}s}(\tau) \: 
e^{A_2 [\eta_{{\bf n}s}(\tau)]} = Z'_{\rm ph} \cdot e^{A_1[\, {\bar \xi}_{{\bf n}s}(\tau)]} \: .
\label{eq:nine}
\end{equation}
The path integral over $\eta(\tau)$ does not depend on any dynamical variables
and therefore can be assigned a multiplicative constant $Z'_{\rm ph}$.  It 
represents the incomplete thermodynamic partition function of freely vibrating
ions. [The partition function will become complete after the final integration
over the end ion coordinates in (\ref{eq:two}).]  The functional $A_1$ in
eq.~(\ref{eq:nine}) has the following form:
\begin{equation}
A_1[{\bar \xi}_{{\bf n}s}(\tau)] = \sum_{{\bf n}s} \left\{ \frac{m_s}{2 \hbar^2} \left[ 
{\dot {\bar \xi}}_{{\bf n}s}(0) {\bar \xi}_{{\bf n}s}(0) - 
{\dot {\bar \xi}}_{{\bf n}s}(\beta) {\bar \xi}_{{\bf n}s}(\beta)
\right] + \frac{1}{2} \int^{\beta}_0 f_{{\bf n}s}(\tau') 
{\bar \xi}_{{\bf n}s}(\tau') {\rm d} \tau'  \right\} \: .
\label{eq:ten}
\end{equation}
As soon as the solution of system (\ref{eq:four}) is known, 
$W$ is calculated explicitly using eqs.~(\ref{eq:nine}) and (\ref{eq:ten}).

\section{ \label{sec:three}
Solution of classical equations
}

In this section we solve the classical equations (\ref{eq:four}) under the 
boundary conditions (\ref{eq:five})-(\ref{eq:six}).  The strategy is to transform
$\xi_{{\bf n}s}$ to normal coordinates, in which the system is diagonal, and then
apply the formula for a single harmonic oscillator.  The first step is a 
Fourier-mass transformation
\begin{equation}
{\bar \xi}_{{\bf n}s}(\tau) = \frac{1}{N} \sum_{\bf k} 
\frac{a_{{\bf k}s}(\tau)}{\sqrt{m_s}} \, e^ {i{\bf k}{\bf n}} \: ,
\label{eq:eleven}
\end{equation}
\begin{equation}
a_{{\bf k}s}(\tau) = 
\sqrt{m_s} \sum_{\bf n} {\bar \xi}_{{\bf n}s}(\tau) e^ {- i {\bf k}{\bf n}} \: ,
\label{eq:twelve}
\end{equation}
where $N$ is the number of unit cells and ${\bf k}$ spans the Brillouin zone of the
reciprocal lattice.  Substituting (\ref{eq:eleven}) into (\ref{eq:four}), multiplying
by $e^{-i {\bf k'}{\bf n}}$ and summing over ${\bf n}$, equation (\ref{eq:four}) is
transformed into the following:
\begin{equation}
- {\ddot a}_{{\bf k}s}(\tau) +  
\hbar^2 \sum_{s'} w_{ss'}({\bf k}) a_{{\bf k}s'}(\tau) = 
\frac{\hbar^2}{\sqrt{m_s}} \sum_{\bf n} e^{-i {\bf k}{\bf n}} f_{{\bf n}s}(\tau) \: ,
\label{eq:thirteen}
\end{equation}
\begin{equation}
w_{ss'}({\bf k}) = w^{\ast}_{s's}({\bf k}) = \sum_{{\bf n} - {\bf n}'} 
\frac{v_{{\bf n}s,{\bf n}'s'}}{\sqrt{m_s m_{s'}}} \, 
e^{i {\bf k}({\bf n}-{\bf n}')} \: .
\label{eq:fourteen}
\end{equation}
The new $p \times p$ matrix $w({\bf k})$ is Hermitian.  The latter property is a 
consequence of the factor $\sqrt{m_s}$ in the above transformations.  This is of 
course a standard device of the classical theory of lattice vibrations \cite{Born}.   
The dimensionality of $w({\bf k})$ is frequency squared.  On the next step, consider
an eigenvalue problem for the matrix $w({\bf k})$:
\begin{equation}
\sum_{s'} w_{ss'}({\bf k}) \, u^{\bf k}_{s'\lambda} = 
\omega^2_{{\bf k} \lambda}    u^{\bf k}_{s \lambda} \: .
\label{eq:fifteen}
\end{equation}
Since $w$ is Hermitian, the eigenvalues $\omega^2_{{\bf k} \lambda}$ are real.  
We also assume that $\omega^2_{{\bf k} \lambda}$ are positive which implies no
structural phase transition.  Further, since 
$w^{\ast}_{ss'}(- {\bf k}) = w_{ss'}({\bf k})$ then 
$\omega_{-{\bf k} \lambda} = \omega_{{\bf k} \lambda}$.  The index $\lambda$ numbers 
different eigenvalues, that is phonon modes.  The eigenvector $u^{\bf k}_{s \lambda}$ 
defines the phonon polarization.  The eigenvectors are orthogonal, 
$\sum_{s} u^{{\bf k}\ast}_{s\lambda} u^{\bf k}_{s \lambda'} = \delta_{\lambda \lambda'}$ .
In ${\bf k}$-points of high symmetry, the eigenvectors can be made orthogonal by a 
proper procedure.  In addition, $u^{-{\bf k} \ast}_{s\lambda} = u^{\bf k}_{s\lambda}$.
Next, expand the functions $a_{{\bf n}s}(\tau)$ in the basis of $u$:
\begin{equation}
a_{{\bf k}s}(\tau) = \sum^{p}_{\lambda = 1} 
c_{{\bf k}\lambda}(\tau) \, u^{\bf k}_{s \lambda} \: ,
\label{eq:seventeen}
\end{equation}
\vspace{-0.5cm}
\begin{equation}
c_{{\bf k}\lambda}(\tau) = \sum^{p}_{s = 1} 
a_{{\bf k}s}(\tau) \, u^{{\bf k}\ast}_{s \lambda} \: ,
\label{eq:eighteen}
\end{equation}
where $c_{{\bf k}\lambda}(\tau)$ are new complex functions of imaginary time.  
Substituting (\ref{eq:seventeen}) into (\ref{eq:thirteen}), using the definition
(\ref{eq:fifteen}), multiplying the resulting equation by 
$u^{{\bf k}\ast}_{s\lambda'}$ and summing over $s$ one obtains
\begin{equation}
- {\ddot c}_{{\bf k}\lambda}(\tau) + 
(\hbar\omega_{{\bf k}\lambda})^2 c_{{\bf k}\lambda}(\tau) = 
\hbar^2 \sum_s \frac{u^{{\bf k}\ast}_{s\lambda}}{\sqrt{m_s}} 
\sum_{\bf n} e^{-i {\bf k}{\bf n}} f_{{\bf n}s}(\tau) \: .
\label{eq:nineteen}
\end{equation}
Thus, the system of equations is fully diagonalized, with a transformed inhomogeneous
term on the right hand side.  The problem is reduced to the well studied case of 
{\em one} harmonic oscillator under an arbitrary external force.  Before applying the 
corresponding formulas the boundary conditions must be recalculated in the new basis.  
Combining eqs.~(\ref{eq:five}), (\ref{eq:six}), (\ref{eq:twelve}), and 
(\ref{eq:eighteen}), the transformed boundary conditions read  
\begin{equation}
c_{{\bf k}\lambda}(0) = \sum_{{\bf n}s} \sqrt{m_s} \, u^{{\bf k}\ast}_{s\lambda}
\, \xi_{{\bf n}s0} \, e^{-i {\bf k}{\bf n}} \: ,
\label{eq:twenty}
\end{equation}
\vspace{-0.2cm}
\begin{equation}
c_{{\bf k}\lambda}(\beta) = \sum_{{\bf n}s} \sqrt{m_s} \, u^{{\bf k}\ast}_{s\lambda}
\, \xi_{{\bf n}s\beta} \, e^{-i {\bf k}{\bf n}} \: .
\label{eq:twentyone}
\end{equation}
To proceed further we recall the following result from the harmonic oscillator theory.  
A {\em real} function $H(\tau)$ satisfying the equation 
\begin{equation}
- {\ddot H}(\tau) + (\hbar\omega_{{\bf k}\lambda})^2 H(\tau) = F(\tau) \: ,
\label{eq:twentytwo}
\end{equation}
with the boundary conditions $H(0) = H_0$ and $H(\beta) = H_{\beta}$, is given by
the expression 
\begin{equation}
H(\tau) = H_0 \,  \frac{\sinh \hbar \omega_{{\bf k}\lambda}(\beta-\tau)}
                       {\sinh \hbar \omega_{{\bf k}\lambda} \beta} +
     H_{\beta} \, \frac{\sinh \hbar \omega_{{\bf k}\lambda} \tau}
                       {\sinh \hbar \omega_{{\bf k}\lambda} \beta} +
\int^{\beta}_0 G_{{\bf k}\lambda}(\tau,\tau') F(\tau') \, {\rm d} \tau' \: ,
\label{eq:twentythree}
\end{equation}
where
\begin{equation}
G_{{\bf k}\lambda}(\tau,\tau') = 
\frac{1}{\hbar \omega_{{\bf k}\lambda} \sinh \hbar \omega_{{\bf k}\lambda} \beta} \left\{ 
\begin{array}{ll}
\sinh \hbar \omega_{{\bf k}\lambda} \tau \cdot 
\sinh \hbar \omega_{{\bf k}\lambda} ( \beta - \tau' ) \: ; & 
\hspace{0.5cm} \tau < \tau' 
\\
\sinh \hbar \omega_{{\bf k}\lambda} ( \beta - \tau ) \cdot 
\sinh \hbar \omega_{{\bf k}\lambda} \tau' \: ; & 
\hspace{0.5cm} \tau > \tau'  
\end{array}
\right.  \: ,
\label{eq:twentyfour}
\end{equation}
is the Green's function of the equation (\ref{eq:twentythree}) under the zero boundary
conditions.  Separating eqs.~(\ref{eq:nineteen})-(\ref{eq:twentyone}) into real and
imaginary parts and applying formula (\ref{eq:twentythree}) one obtains
\begin{eqnarray}
c_{{\bf k}\lambda}(\tau) & = & \frac{\sinh \hbar \omega_{{\bf k}\lambda}(\beta-\tau)}
                                    {\sinh \hbar \omega_{{\bf k}\lambda} \beta} 
\sum_{{\bf n}s} \sqrt{m_s} \, u^{{\bf k}\ast}_{s\lambda} \, 
\xi_{{\bf n}s0} \, e^{-i {\bf k}{\bf n}} +
                               \frac{\sinh \hbar \omega_{{\bf k}\lambda} \tau}
                                    {\sinh \hbar \omega_{{\bf k}\lambda} \beta} 
\sum_{{\bf n}s} \sqrt{m_s} \, u^{{\bf k}\ast}_{s\lambda} \, 
\xi_{{\bf n}s\beta} \, e^{-i {\bf k}{\bf n}}                 \nonumber \\
        &  & + 
\int^{\beta}_0 {\rm d} \tau' G_{{\bf k}\lambda}(\tau,\tau') 
\, \hbar^2 \sum_s \frac{u^{{\bf k}\ast}_{s\lambda}}{\sqrt{m_s}} 
\sum_{\bf n} e^{-i {\bf k}{\bf n}} f_{{\bf n}s}(\tau') \: .
\label{eq:twentyfive}
\end{eqnarray}
Restoration of the classical displacement ${\bar \xi}_{{\bf n}s}$ from eq.~(\ref{eq:eleven})
yields the complete solution of the original Cauchy problem (\ref{eq:four})-(\ref{eq:six})
\begin{eqnarray}
{\bar \xi}_{{\bf n}s}(\tau) & = & 
\frac{1}{N} \sum_{{\bf k}\lambda} \frac{\sinh \hbar \omega_{{\bf k}\lambda}(\beta-\tau)}
                                       {\sinh \hbar \omega_{{\bf k}\lambda} \beta} 
\sum_{{\bf n'}s'} \sqrt{\frac{m_{s'}}{m_s}} \, 
u^{{\bf k}\ast}_{s'\lambda} u^{\bf k}_{s\lambda} \, 
 e^{i {\bf k}({\bf n}-{\bf n'})} \, \xi_{{\bf n'}s'0}  \nonumber \\
& & +                                
\frac{1}{N} \sum_{{\bf k}\lambda} \frac{\sinh \hbar \omega_{{\bf k}\lambda} \tau}
                                       {\sinh \hbar \omega_{{\bf k}\lambda} \beta} 
\sum_{{\bf n'}s'} \sqrt{\frac{m_{s'}}{m_s}} \, 
u^{{\bf k}\ast}_{s'\lambda} u^{\bf k}_{s\lambda} \, 
 e^{i {\bf k}({\bf n}-{\bf n'})} \, \xi_{{\bf n'}s'\beta}  \nonumber \\
        &  & + 
\frac{1}{N} \sum_{{\bf k}\lambda}
\int^{\beta}_0 {\rm d} \tau' G_{{\bf k}\lambda}(\tau,\tau') \,
\sum_{{\bf n'}s'} \frac{\hbar^2}{\sqrt{m_s m_{s'}}} \, 
u^{{\bf k}\ast}_{s'\lambda} u^{\bf k}_{s\lambda} \, 
 e^{i {\bf k}({\bf n}-{\bf n'})} \, f_{{\bf n'}s'}(\tau')  \: .
\label{eq:twentysix}
\end{eqnarray}
Note that because
\begin{equation}
u^{-{\bf k}\ast}_{s'\lambda} u^{-{\bf k}}_{s\lambda} \, e^{- i {\bf k}({\bf n}-{\bf n'})} 
= u^{\bf k}_{s'\lambda} u^{{\bf k}\ast}_{s\lambda} \, e^{- i {\bf k}({\bf n}-{\bf n'})}
= \left[  u^{{\bf k}\ast}_{s'\lambda} u^{\bf k}_{s\lambda} \, 
 e^{i {\bf k}({\bf n}-{\bf n'})} \right]^{\ast} \: ,
\label{eq:twentyseven}
\end{equation}
the summation over ${\bf k}$ in equation (\ref{eq:twentysix}) makes the displacements 
purely real, as expected.  Finally, substituting the classical solution in 
eq.~(\ref{eq:ten}), after some algebra, one obtains the polaron action $A_1$ 
as a function of the end ion coordinates and a functional of the forces
\begin{eqnarray}
& & \hspace{-0.4cm}
A_1 [\xi_{{\bf n}s0}, \xi_{{\bf n}s\beta}; f_{{\bf n}s}(\tau) ] = \nonumber \\ 
& & \hspace{-0.3cm} \frac{1}{2N} \sum_{{\bf k}\lambda} 
\frac{\omega_{{\bf k}\lambda}}{\hbar \sinh \hbar \omega_{{\bf k}\lambda} \beta} 
\sum_{{\bf n}s {\bf n'}s'} \!\! \sqrt{m_{s'} m_s} \,\, 
\Re \left[ u^{{\bf k}\ast}_{s'\lambda} u^{\bf k}_{s\lambda} \, 
e^{i {\bf k}({\bf n}-{\bf n'})} \right]  \times
\nonumber \\
& & \times \left\{ - (\xi_{{\bf n}s0} \xi_{{\bf n'}s'0} + 
\xi_{{\bf n}s\beta} \xi_{{\bf n'}s'\beta}) \cosh{\hbar \omega_{{\bf k}\lambda} \beta}
+ \xi_{{\bf n}s0} \xi_{{\bf n'}s'\beta} + \xi_{{\bf n}s\beta} \xi_{{\bf n'}s'0}
\right\}                                                              \nonumber \\
& & + \frac{1}{2N} \sum_{{\bf k}\lambda} \sum_{{\bf n}s {\bf n'}s'}  
\Re \left[ u^{{\bf k}\ast}_{s'\lambda} u^{\bf k}_{s\lambda} \, 
e^{i {\bf k}({\bf n}-{\bf n'})} \right] 
\int^{\beta}_0 {\rm d} \tau' \frac{\sinh \hbar \omega_{{\bf k}\lambda} (\beta - \tau')}
                                  {\sinh \hbar \omega_{{\bf k}\lambda} \beta} 
\left\{ \sqrt{\frac{m_s}{m_{s'}}} \xi_{{\bf n}s0} f_{{\bf n}'s'}(\tau') 
      + \sqrt{\frac{m_{s'}}{m_s}} \xi_{{\bf n}'s'0} f_{{\bf n}s}(\tau') \right\} \nonumber \\
& & + \frac{1}{2N} \sum_{{\bf k}\lambda} \sum_{{\bf n}s {\bf n'}s'}  
\Re \left[ u^{{\bf k}\ast}_{s'\lambda} u^{\bf k}_{s\lambda} \, 
e^{i {\bf k}({\bf n}-{\bf n'})} \right] 
\int^{\beta}_0 {\rm d} \tau' \frac{\sinh \hbar \omega_{{\bf k}\lambda} \tau'}
                                  {\sinh \hbar \omega_{{\bf k}\lambda} \beta} 
\left\{ \sqrt{\frac{m_s}{m_{s'}}} \xi_{{\bf n}s\beta} f_{{\bf n}'s'}(\tau') 
      + \sqrt{\frac{m_{s'}}{m_s}} \xi_{{\bf n}'s'\beta} f_{{\bf n}s}(\tau') \right\} \nonumber \\
& & + \frac{1}{2N} \sum_{{\bf k}\lambda} \sum_{{\bf n}s {\bf n'}s'}
\frac{\hbar^2}{\sqrt{m_s m_{s'}}} \, 
\Re \left[ u^{{\bf k}\ast}_{s'\lambda} u^{\bf k}_{s\lambda} \, 
e^{i {\bf k}({\bf n}-{\bf n'})} \right]  
 \int^{\beta}_0 \!\!\! \int^{\beta}_0 {\rm d} \tau {\rm d} \tau' 
G_{{\bf k}\lambda}(\tau,\tau') f_{{\bf n}s}(\tau) f_{{\bf n'}s'}(\tau')  \: .
\label{eq:twentyeight}
\end{eqnarray}
It should be noted that the real part symbol $\Re$ is optional here, the reality of
the action is guarantied by summation over ${\bf k}$ and the above-mentioned symmetry
properties of the integrand.  Equations (\ref{eq:nine}) and (\ref{eq:twentyeight})
provide an explicit solution for the quantity $W$ defined in eq.~(\ref{eq:three}).

\section{ \label{sec:four}
Polaron action
}

In accordance with the general ideas of the continuous time polaron Monte Carlo,
the end points in the action (\ref{eq:twentyeight}) must be related by 
$\xi_{{\bf n}s\beta} = \xi_{{\bf n}-\Delta {\bf r} s \, 0}$. 
Then $W = Z'_{\rm ph} \cdot e^{A_1}$ should be integrated over all 
$\xi_{{\bf n}s0}$ from minus to plus infinity, see eqs.~(\ref{eq:two}), 
(\ref{eq:three}), and (\ref{eq:nine}).  To perform the integration, it is 
convenient to employ the same series of transformations that enabled calculation 
of the path integral, that is 
\begin{equation}
\xi_{{\bf n}s0} = \frac{1}{N} \sum_{{\bf k}\lambda} \frac{1}{\sqrt{m_s}} \, 
u^{\bf k}_{s\lambda} e^{i {\bf k} {\bf n}} g_{{\bf k}\lambda} = 
\frac{1}{N} \sum_{{\bf k}\lambda} \frac{1}{\sqrt{m_s}} \, 
u^{{\bf k} \ast}_{s\lambda} e^{- i {\bf k} {\bf n}} g^{\ast}_{{\bf k}\lambda} \: ,
\label{eq:thirty}
\end{equation}
\begin{equation}
\xi_{{\bf n}s\beta} = \frac{1}{N} \sum_{{\bf k}\lambda} \frac{1}{\sqrt{m_s}} \, 
u^{\bf k}_{s\lambda} e^{i {\bf k} ({\bf n} - \Delta {\bf r})} g_{{\bf k}\lambda} = 
\frac{1}{N} \sum_{{\bf k}\lambda} \frac{1}{\sqrt{m_s}} \, 
u^{{\bf k} \ast}_{s\lambda} e^{- i {\bf k} ({\bf n} - \Delta {\bf r})} 
g^{\ast}_{{\bf k}\lambda} \: .
\label{eq:thirtyone}
\end{equation}
Here $g_{{\bf k}\lambda} = b_{{\bf k}\lambda} + i d_{{\bf k}\lambda}$ is a new
complex variable.  Substituting the expansions into eq.~(\ref{eq:twentyeight}) 
one obtains the action as a function of amplitudes $b$ and $d$
\begin{eqnarray}
A_1 [b_{{\bf k}\lambda}, d_{{\bf k}\lambda}; f_{{\bf n}s}(\tau)] 
& = & \frac{1}{N} \sum_{{\bf k}\lambda} 
\frac{\omega_{{\bf k}\lambda}}{\hbar \sinh \hbar \omega_{{\bf k}\lambda} \beta}
\left( \cosh \hbar \omega_{{\bf k}\lambda} \beta - \cos {\bf k} \Delta {\bf r} \right)
\left( - b^2_{{\bf k}\lambda} - d^2_{{\bf k}\lambda} \right)   \nonumber \\
& + & \frac{1}{N} \sum_{{\bf k}\lambda} b_{{\bf k}\lambda} 
\sum_{{\bf n}s} \frac{1}{\sqrt{m_s}} 
\left\{ \Re \left[ u^{\bf k}_{s\lambda} e^{i {\bf k}{\bf n}} \right] 
B_{{\bf k} \lambda {\bf n} s}
      + \Re \left[ u^{\bf k}_{s\lambda} e^{i {\bf k}({\bf n}-\Delta {\bf r})} \right] 
C_{{\bf k} \lambda {\bf n} s}     \right\}                     \nonumber \\
& - & \frac{1}{N} \sum_{{\bf k}\lambda} d_{{\bf k}\lambda} 
\sum_{{\bf n}s} \frac{1}{\sqrt{m_s}} 
\left\{ \Im \left[ u^{\bf k}_{s\lambda} e^{i {\bf k}{\bf n}} \right] 
B_{{\bf k} \lambda {\bf n} s}
      + \Im \left[ u^{\bf k}_{s\lambda} e^{i {\bf k}({\bf n}-\Delta {\bf r})} \right] 
C_{{\bf k} \lambda {\bf n} s}     \right\}                     \nonumber \\
& + & \frac{1}{2N} \sum_{{\bf k}\lambda} \sum_{{\bf n}s {\bf n'}s'}
\frac{\hbar^2}{\sqrt{m_s m_{s'}}} \, 
\Re \left[ u^{{\bf k}\ast}_{s'\lambda} u^{\bf k}_{s\lambda} \, 
e^{i {\bf k}({\bf n}-{\bf n'})} \right]  
 \int^{\beta}_0 \!\!\! \int^{\beta}_0 {\rm d} \tau {\rm d} \tau' 
G_{{\bf k}\lambda}(\tau,\tau') f_{{\bf n}s}(\tau) f_{{\bf n'}s'}(\tau')  \: ,
\label{eq:thirtytwo}
\end{eqnarray}
where
\begin{equation}
B_{{\bf k} \lambda {\bf n} s} \equiv
\int^{\beta}_0 {\rm d} \tau' \frac{\sinh \hbar \omega_{{\bf k}\lambda} (\beta-\tau')}
                                  {\sinh \hbar \omega_{{\bf k}\lambda} \beta}
f_{{\bf n}s} (\tau')  \: , 
\label{eq:thirtythree}
\end{equation}
\vspace{-0.5cm}
\begin{equation}
C_{{\bf k} \lambda {\bf n} s} \equiv
\int^{\beta}_0 {\rm d} \tau' \frac{\sinh \hbar \omega_{{\bf k}\lambda} \tau'}
                                  {\sinh \hbar \omega_{{\bf k}\lambda} \beta}
f_{{\bf n}s} (\tau')  \: , 
\label{eq:thirtyfour}
\end{equation}
and $\Re$ and $\Im$ denote the real and imaginary parts, respectively. 
The final step is gaussian integration of $e^{A_1}$ over the real $b$ and $d$ from
minus to plus infinity.  The reality of $\xi$ and the property 
$u^{-{\bf k} \ast}_{s\lambda} = u^{\bf k}_{s\lambda}$ ensure that 
$b_{-{\bf k}\lambda} = b_{{\bf k}\lambda}$ and $d_{-{\bf k}\lambda} = - d_{{\bf k}\lambda}$.
Thus the domain of integration is limited to half of the Brillouin zone.  
The integration is straightforward, and leads to the result
$const \cdot e^{A_{\Delta {\bf r}}}$ where $const$ completes the partition
function of free phonons while the polaron action $A_{\Delta {\bf r}}$ is given by 
\begin{eqnarray}
A_{\Delta {\bf r}} [f_{{\bf n}s}(\tau)] & = & \frac{1}{N} \sum_{{\bf k}\lambda}
\frac{\hbar \sinh \hbar \omega_{{\bf k}\lambda} \beta} {4\omega_{{\bf k}\lambda}
\left( \cosh \hbar \omega_{{\bf k}\lambda} \beta - \cos {\bf k} \Delta {\bf r} \right) }
\sum_{{\bf n} s {\bf n}' s'} \frac{1}{\sqrt{m_s m_{s'}}}  \times       \nonumber \\
& \times & \left\{ \Re \left[ u^{\bf k}_{s\lambda} e^{i {\bf k}{\bf n}} 
u^{{\bf k}\ast}_{s'\lambda} e^{-i {\bf k}{\bf n}'} \right] 
B_{{\bf k} \lambda {\bf n} s} B_{{\bf k} \lambda {\bf n}' s'}
+ \Re \left[ u^{\bf k}_{s\lambda} e^{i {\bf k}{\bf n}} 
u^{{\bf k}\ast}_{s'\lambda} e^{-i {\bf k}({\bf n}' - \Delta {\bf r})} \right] 
B_{{\bf k} \lambda {\bf n} s} C_{{\bf k} \lambda {\bf n}' s'} \right.   \nonumber \\
& & + \left. \Re \left[ u^{\bf k}_{s\lambda} e^{i {\bf k}({\bf n}-\Delta {\bf r})} 
u^{{\bf k}\ast}_{s'\lambda} e^{-i {\bf k}{\bf n}'} \right] 
C_{{\bf k} \lambda {\bf n} s} B_{{\bf k} \lambda {\bf n}' s'}
+ \Re \left[ u^{\bf k}_{s\lambda} e^{i {\bf k}({\bf n} - \Delta {\bf r})} 
u^{{\bf k}\ast}_{s'\lambda} e^{-i {\bf k}({\bf n}' - \Delta {\bf r})} \right] 
C_{{\bf k} \lambda {\bf n} s} C_{{\bf k} \lambda {\bf n}' s'} \right\}   \nonumber \\
& + & \frac{1}{2N} \sum_{{\bf k}\lambda} \sum_{{\bf n}s {\bf n'}s'}
\frac{\hbar^2}{\sqrt{m_s m_{s'}}} \, 
\Re \left[ u^{{\bf k}\ast}_{s'\lambda} u^{\bf k}_{s\lambda} \, 
e^{i {\bf k}({\bf n}-{\bf n'})} \right]  
 \int^{\beta}_0 \!\!\! \int^{\beta}_0 {\rm d} \tau {\rm d} \tau' 
G_{{\bf k}\lambda}(\tau,\tau') f_{{\bf n}s}(\tau) f_{{\bf n'}s'}(\tau')  \: .
\label{eq:thirtyfive}
\end{eqnarray}
To simplify this expression consider first the case $\Delta {\bf r} = 0$.  This corresponds
of course to the periodic boundary conditions in imaginary time.  Taking into account
the explicit forms of $B$, $C$, and $G$, the periodic action can be brought into the form
\begin{equation}
A_{0} [f_{{\bf n}s}(\tau)] = \frac{1}{N} \sum_{{\bf k}\lambda}
\frac{\hbar}{4\omega_{{\bf k}\lambda}}
\int^{\beta}_0 \!\!\! \int^{\beta}_0 {\rm d} \tau {\rm d} \tau' 
\frac{\cosh \hbar \omega_{{\bf k}\lambda} ( \frac{\beta}{2} - \vert \tau - \tau' \vert ) } 
{\sinh \hbar \omega_{{\bf k}\lambda} \frac{\beta}{2}}
\sum_{{\bf n} s {\bf n}' s'} \frac{1}{\sqrt{m_s m_{s'}}}
\Re \left[ u^{\bf k}_{s\lambda} u^{{\bf k}\ast}_{s'\lambda} 
e^{i {\bf k}({\bf n} - {\bf n}')} \right]
f_{{\bf n}s}(\tau) f_{{\bf n'}s'}(\tau')  \: .
\label{eq:thirtysix}
\end{equation}
Comparing this result with an analogous expression for independent oscillators 
\cite{Feynman1972} shows that eq.~(\ref{eq:thirtysix}) involves the forces 
transformed to the basis in which the phonon subsystems is diagonal.  This is of 
course exactly what is expected intuitively.  

Returning to the general action (\ref{eq:thirtyfive}), one can calculate the correction
to the periodic action by taking the difference between $A_{\Delta {\bf r}}$ and $A_0$.
The general expression is rather complicated.  However, nonzero $\Delta {\bf r}$ are
needed only in calculation of the polaron mass and spectrum, which are the properties
of zero temperature partition functions.  Thus the action correction is needed 
only in the $\beta \rightarrow \infty$ limit.  Then $\cos{{\bf k} \Delta {\bf r}}$
is negligible compared to $\cosh \hbar \omega_{{\bf k}\lambda} \beta$, and 
$\tanh \hbar \omega_{{\bf k}\lambda} \beta \approx 1$ with exponential accuracy. 
Thus the correction becomes
\begin{eqnarray}
\Delta A_{\Delta {\bf r}} (\beta \rightarrow \infty) & = &
\frac{1}{N} \sum_{{\bf k}\lambda} \frac{\hbar}{4\omega_{{\bf k}\lambda}}
\sum_{{\bf n} s {\bf n}' s'} \frac{1}{\sqrt{m_s m_{s'}}}
\left\{ \Re \left[ u^{\bf k}_{s\lambda} u^{{\bf k}\ast}_{s'\lambda} 
\left( e^{i {\bf k}({\bf n} - {\bf n}' + \Delta {\bf r})} - 
e^{i {\bf k}({\bf n} - {\bf n}')} \right)  \right]
B_{{\bf k} \lambda {\bf n} s} C_{{\bf k} \lambda {\bf n}' s'} +  \right. \nonumber \\
& & \makebox[2.0cm]{} \left. + \Re \left[ u^{\bf k}_{s\lambda} u^{{\bf k}\ast}_{s'\lambda} 
\left( e^{i {\bf k}({\bf n} - {\bf n}' - \Delta {\bf r})} - 
e^{i {\bf k}({\bf n} - {\bf n}')} \right)  \right]
C_{{\bf k} \lambda {\bf n} s} B_{{\bf k} \lambda {\bf n}' s'}  \right\} = 
  \nonumber \\
& = & \frac{1}{N} \sum_{{\bf k}\lambda} \frac{\hbar}{2\omega_{{\bf k}\lambda}}
\sum_{{\bf n} s {\bf n}' s'} \frac{1}{\sqrt{m_s m_{s'}}}
\Re \left[ u^{\bf k}_{s\lambda} u^{{\bf k}\ast}_{s'\lambda} 
\left( e^{i {\bf k}({\bf n} - {\bf n}' + \Delta {\bf r})} - 
e^{i {\bf k}({\bf n} - {\bf n}')} \right)  \right]
B_{{\bf k} \lambda {\bf n} s} C_{{\bf k} \lambda {\bf n}' s'} \: .
\label{eq:thirtyseven}
\end{eqnarray}

To summarize the results, the shift partition function (\ref{eq:two}) reduces to a 
path integral over only the electron coordinates
\begin{equation}
Z_{\Delta {\bf r}} = Z_{\rm ph} \sum_{{\bf r}_0} 
\int^{{\bf r}(\beta) = {\bf r}_0 + \Delta {\bf r}}_{{\bf r}(0) = {\bf r}_0} 
{\cal D} {\bf r}(\tau) \: 
e^{T[ \, {\bf r}(\tau)] + A_0 [ \, {\bf r}(\tau) ] + 
\Delta A_{\Delta {\bf r}} [ \, {\bf r}(\tau) ] } \: .
\label{eq:thirtyeight}
\end{equation}
Here 
$Z_{\rm ph} = \prod_{{\bf k}\lambda} [2 \sinh (\frac{1}{2} \hbar \beta \omega_{{\bf k}\lambda})]^{-1}$ 
is the full partition of free phonons.  It is a multiplicative
constant that does not influence polaron dynamics and drops out of the configuration
weight in Monte Carlo simulations.  $T$ is the action piece originating from the 
kinetic energy of the particle.  On the lattice and in continuous imaginary time 
treatment of the kinetic energy is non-trivial as has been described elsewhere
\cite{Kornilovitch1998}.  The periodic action $A_0$ is given by the formula 
(\ref{eq:thirtysix}) which is valid for any temperature.  This action can be used 
to calculate equilibrium polaron thermodynamics: free energy, specific heat, static 
correlation functions, and so on.  The action correction $\Delta A$ is due to shifted 
boundary conditions (non-zero $\Delta {\bf r}$) and is given by equations 
(\ref{eq:thirtyseven}) and (\ref{eq:thirtythree})-(\ref{eq:thirtyfour}).  
The formulae are valid in the low temperature limit, i.e. the difference between 
eq.~(\ref{eq:thirtyseven}) and the exact formula is 
${\cal O}(e^{- \hbar \omega_{{\bf k}\lambda} \beta})$. 
Equation (\ref{eq:thirtyseven}) can be used to calculate the polaron mass, spectrum,
and density of states.  If for some reason the full $\Delta {\bf r} \neq 0$ action
is needed at a finite temperature, then the exact expression (\ref{eq:thirtyfive})
should be used.

\section{ \label{sec:five}
Special cases
}   

The formulae derived in the preceeding section are general but can be difficult to 
evaluate.  In this section, simplified expressions for several special cases are 
obtained that can be used in practical calculations.

\subsection{ \label{sec:fiveone}
Independent oscillators
}   

The independence of oscillators implies that the force constants are diagonal in 
the unit cell index and in the degree of freedom index: 
$v_{{\bf n}s, {\bf n}'s'} = v_s \delta_{{\bf n} {\bf n}'} \delta_{ss'}$.  Then the
matrix elements $w_{ss'}({\bf k}) = (v_s/m_s) \delta_{ss'}$ are real and independent of 
the momentum, see eq.~(\ref{eq:fourteen}).  The eigenvalue problem (\ref{eq:fifteen}) 
becomes diagonal with eigenvectors $u_{s\lambda} = \delta_{s\lambda}$ and
eigenvalues $\omega^2_{\lambda} = v_{\lambda}/m_{\lambda}$, where $1 \leq \lambda \leq p$.  
Neither $u$ nor $\omega$ depend on ${\bf k}$.  Substitution into the periodic action
(\ref{eq:thirtysix}) leads to the following simplifications: (i) Summation over ${\bf k}$
yields $N \delta_{{\bf n}{\bf n}'}$; (ii) The emerged delta function removes the summation
over ${\bf n}'$; (iii) Summations over $s$ and $s'$ are performed using the explicit form
of the eigenvectors $u$; (iv) The summation index $\lambda$ is changed back to index
$s$ to unify the notation.  As a result, the periodic function takes the form:
\begin{equation}
A_{0} [f_{{\bf n}s}(\tau)] = \sum^{p}_{s = 1} \sum_{\bf n}
\frac{\hbar}{4\omega_{s} m_{s}}
\int^{\beta}_0 \!\!\! \int^{\beta}_0 {\rm d} \tau {\rm d} \tau' 
\frac{\cosh \hbar \omega_{s} ( \frac{\beta}{2} - \vert \tau - \tau' \vert ) } 
{\sinh \hbar \omega_{s} \frac{\beta}{2}} f_{{\bf n}s}(\tau) f_{{\bf n}s}(\tau')  \: ,
\label{eq:thirtynine}
\end{equation}
where $\omega_s = \sqrt{v_s/m_s}$.  After the same transformations, the action correction 
(\ref{eq:thirtyseven}) becomes  
\begin{equation}
\Delta A [f_{{\bf n}s}(\tau)] = \sum^{p}_{s = 1} \sum_{\bf n}
\frac{\hbar}{2\omega_{s} m_{s}} B_{{\bf n}s} 
\left( C_{{\bf n} + \Delta {\bf r}, s} - C_{{\bf n}s} \right)  \: ,
\label{eq:forty}
\end{equation}
\begin{equation}
B_{{\bf n} s} \equiv
\int^{\beta}_0 {\rm d} \tau' \frac{\sinh \hbar \omega_{s} (\beta-\tau')}
                     {\sinh \hbar \omega_{s} \beta} f_{{\bf n}s} (\tau')  \: , 
\label{eq:fortyone}
\end{equation}
\vspace{-0.2cm}
\begin{equation}
C_{{\bf n} s} \equiv
\int^{\beta}_0 {\rm d} \tau' \frac{\sinh \hbar \omega_{s} \tau'}
                     {\sinh \hbar \omega_{s} \beta} f_{{\bf n}s} (\tau')  \: . 
\label{eq:fortytwo}
\end{equation}
These expressions can be used in analysis of isotropic and anisotropic spherical 
oscillators in two, three, and higher dimensions.

\subsection{ \label{sec:fivetwo}
One phonon branch
}   

Substantial simplifications take place in the case of only one phonon degree of freedom
per unit cell with mass $m$ and force function $v({\bf n}-{\bf n}')$.  It should be 
emphasized that this does not restrict the dimensionality of the lattice in any way.  
Rather, the electron interacts predominantly with one type of ion displacement.  
Mathematically, the indices $s$ and $\lambda$ accept only one value 1, and therefore 
can be omitted hereafter.  The eigenvalue matrix equation (\ref{eq:fifteen}) reduces 
to one linear equation with the solution $u^{\bf k} = 1$ and
\begin{equation}
\omega^2_{\bf k} = w({\bf k}) = \frac{1}{m} \sum_{\bf n} v({\bf n}) \,
e^{i{\bf k}{\bf n}}  \: . 
\label{eq:fortythree}
\end{equation}
Thus, the phonon frequency is explicitly known.  Note that $\omega^2_{\bf k}$ is 
automatically real because $v(-{\bf n}) = v({\bf n})$.  These results transform the action 
as follows
\begin{equation}
A_{0} [f_{{\bf n}s}(\tau)] = \frac{1}{N} \sum_{\bf k}
\frac{\hbar}{4 m \omega_{\bf k}}
\int^{\beta}_0 \!\!\! \int^{\beta}_0 {\rm d} \tau {\rm d} \tau' 
\frac{\cosh \hbar \omega_{\bf k} ( \frac{\beta}{2} - \vert \tau - \tau' \vert ) } 
{\sinh \hbar \omega_{\bf k} \frac{\beta}{2}}
\sum_{{\bf n} {\bf n}'} \cos [{\bf k}({\bf n} - {\bf n}')]
f_{{\bf n}}(\tau) f_{{\bf n'}}(\tau')  \: ,
\label{eq:fortyfour}
\end{equation}
\begin{equation}
\Delta A [f_{{\bf n}s}(\tau)] = \frac{1}{N} \sum_{\bf k}
\frac{\hbar}{2 m \omega_{\bf k}} \sum_{{\bf n} {\bf n}'} 
\left\{ \cos [{\bf k}({\bf n} - {\bf n}' + \Delta {\bf r})] - 
       \cos [{\bf k}({\bf n} - {\bf n}')] \right\}
B_{{\bf k}{\bf n}} C_{{\bf k}{\bf n'}}  \: ,
\label{eq:fortyfive}
\end{equation}
where $B$ and $C$ are given by (\ref{eq:thirtythree})-(\ref{eq:thirtyfour}) with 
indices $s$ and $\lambda$ removed.  The derived action should be used, e.g., in 
analysis of the polaron on a linear chain of one-dimensional coupled oscillators
\cite{DeRaedt1984}.

\subsection{ \label{sec:fivethree}
Holstein interaction
}   

The Holstein local interaction is one of the most popular polaron models 
\cite{Holstein1959}. At the same time, it is usually studies in combination with an 
Einstein set of decoupled one-dimensional oscillators.  Here we will derive the 
action for the Holstein polaron interacting with an arbitrary phonon subsystem.  
We assume only one electron Wannier state $\vert {\bf r} \rangle$ per unit cell, 
leaving the multi-orbital case for future work.  In the spirit of the Holstein 
model, we further assume that the electron interacts only with the ion degrees 
of freedom of the unit cell it currently occupies, i.e.
\begin{equation}
f_{{\bf n}s}(\tau) = \kappa_s \delta_{{\bf n} {\bf r}(\tau)} \: ,
\label{eq:fortysix}
\end{equation}
where ${\bf r}(\tau)$ is the electron coordinate at imaginary time $\tau$.  Note that 
the local force constants $\kappa_s$ are {\em different} for different displacements.
The locality of the interaction eliminates the summations over ${\bf n}$ and ${\bf n}'$
in equations (\ref{eq:thirtysix})-(\ref{eq:thirtyseven}).  As a result, the Holstein polaron 
action becomes
\begin{equation}
A_{0} [{\bf r}(\tau)] = \frac{1}{N} \sum_{{\bf k}\lambda}
\frac{\hbar}{4\omega_{{\bf k}\lambda}}
\int^{\beta}_0 \!\!\! \int^{\beta}_0 {\rm d} \tau {\rm d} \tau' 
\frac{\cosh \hbar \omega_{{\bf k}\lambda} ( \frac{\beta}{2} - \vert \tau - \tau' \vert ) } 
{\sinh \hbar \omega_{{\bf k}\lambda} \frac{\beta}{2}}
\sum_{ss'} \frac{\kappa_s \kappa_{s'}}{\sqrt{m_s m_{s'}}}
\Re \left[ u^{\bf k}_{s\lambda} u^{{\bf k}\ast}_{s'\lambda} 
e^{i {\bf k}[ \, {\bf r}(\tau) - \, {\bf r}(\tau')]} \right] \: ,
\label{eq:fortyseven}
\end{equation}
\begin{eqnarray}
\Delta A_{\Delta {\bf r}} [{\bf r}(\tau)] & = & \frac{1}{N}  
\sum_{{\bf k}\lambda} \frac{\hbar}{2\omega_{{\bf k}\lambda}}
\int^{\beta}_0 \!\!\! \int^{\beta}_0 {\rm d} \tau {\rm d} \tau'
\frac{\sinh \hbar \omega_{{\bf k}\lambda} (\beta-\tau) 
      \sinh \hbar \omega_{{\bf k}\lambda} \tau' }
     {\sinh^2 \hbar \omega_{{\bf k}\lambda} \beta}       \times \nonumber \\     
& \times & \sum_{ss'} \frac{\kappa_s \kappa_{s'}}{\sqrt{m_s m_{s'}}} \:
\Re \left[ u^{\bf k}_{s\lambda} u^{{\bf k}\ast}_{s'\lambda} 
\left( e^{i {\bf k}[ \, {\bf r}(\tau) - \, {\bf r}(\tau') + \Delta {\bf r}]} - 
e^{i {\bf k}[ \, {\bf r}(\tau) - \, {\bf r}(\tau')]} \right)  \right] \: .
\label{eq:fortyeight}
\end{eqnarray}

\vspace{0.5cm}

The three cases considered in this section are important on their own right,
but they also serve as examples of how particular models are derived from the
general polaron action.  Many more models can be developed.  In particular, 
one can envision combinations of the above cases: Holstein interaction with
one dispersive phonon branch, Holstein interaction with uncoupled 
three-dimensional oscillators, and so on.

\section{ \label{sec:six}
Discussion
}

In this paper, the polaron action for the general dispersive phonon system has 
been derived.  It is important to list the conditions under which these results 
have been obtained.  First of all, the crystal lattice is treated in the quadratic 
approximation, i.e. the phonons do not interact with each other.  Secondly, the 
electron-phonon interaction is limited to the electron density - ion displacement 
type.  There are other important classes of electron-phonon interactions -- most 
notably the Jahn-Teller and Su-Schrieffer-Heeger ones -- which require separate 
analysis.  Finally, the complexity of the electron kinetic energy has not been 
fully addressed.  This generalization is yet to be addressed in the polaron 
literature.  

The motivation for the present work was the extension of the capabilities of the 
path-integral polaron Monte Carlo \cite{Kornilovitch1998}.  However, the practical 
usefulness of the developed formalism entirely depends on the efficient calculation 
of the action.  Consider the general formulas (\ref{eq:thirtysix}) and 
(\ref{eq:thirtyseven}).  Not only they involve double integration over the imaginary
time, but also summation over all phonon momenta and branches, and a double sum 
over the unit cells and ion degrees of freedom.  Thus direct calculation of the
ful action at every Monte Carlo update may be prohibitively expensive.  Development
of the practical ways to apply the present formalism to specific polaron models
will be the subject of future investigations.

\acknowledgments

The author thanks Sasha Alexandrov, James Hague and John Samson for useful discussions
on the subject of this paper.  The work was supported by EPSRC (UK), grant EP/C518365/1.

\end{document}